\RequirePackage[2020-02-02]{latexrelease}
\documentclass[aps,prb,twocolumn,superscriptaddress,showpacs,aps,longbibliography]{revtex4-2}

\pdfoutput=1
\usepackage[titletoc,toc,title]{appendix}
\usepackage{amsmath,amssymb}
\usepackage{graphicx}
\usepackage{color}
\usepackage{rotating}
\usepackage{bm}
\usepackage{setspace}
\usepackage{hyperref}
\usepackage{float}
\usepackage{soul}
\usepackage[T1]{fontenc}
\usepackage{bbold}
\usepackage{color}
\usepackage{braket}
\usepackage{physics}
\usepackage{subcaption}
\usepackage{scrextend}

\graphicspath{{figures/}}

\hypersetup{
colorlinks=true,
urlcolor= blue,
citecolor=blue,
linkcolor= blue,
bookmarks=true,
bookmarksopen=false,
}

\renewcommand{\vec}[1]{\bm{#1}}

\begin{document}

\title{HubbardNet: Efficient Predictions of the Bose-Hubbard Model Spectrum with Deep Neural Networks}

\author{Ziyan Zhu}
\altaffiliation{Present address: Stanford Institute for Materials and Energy Sciences,
SLAC National Accelerator Laboratory, Menlo Park, CA 94025, USA}
\email{ziyanzhu@stanford.edu}
\affiliation{Department of Physics, Harvard University, Cambridge, Massachusetts 02138, USA}
\author{Marios Mattheakis}
\affiliation{John A. Paulson School of Engineering and Applied Sciences, Harvard University, Cambridge, Massachusetts 02138, USA}
\author{Weiwei Pan}
\affiliation{John A. Paulson School of Engineering and Applied Sciences, Harvard University, Cambridge, Massachusetts 02138, USA}
\author{Efthimios Kaxiras}
\affiliation{Department of Physics, Harvard University, Cambridge, Massachusetts 02138, USA}
\affiliation{John A. Paulson School of Engineering and Applied Sciences, Harvard University, Cambridge, Massachusetts 02138, USA}

\begin{abstract}
We present a deep neural network (DNN)-based model (\texttt{HubbardNet}) to variationally find the ground state and excited state wavefunctions of the one-dimensional and two-dimensional Bose-Hubbard model. Using this model for a square lattice with $M$ sites, we obtain the energy spectrum as an analytical function of the on-site Coulomb repulsion, $U$, and the total number of particles, $N$, from a single training. This approach bypasses the need to solve a new hamiltonian for each different set of values $(U,N)$. Using \texttt{HubbardNet}, we identify the two ground state phases of the Bose-Hubbard model (Mott insulator and superfluid). We show that the DNN-parametrized solutions are in excellent agreement with results from the exact diagonalization of the hamiltonian, and it outperforms exact diagonalization in terms of computational scaling. These advantages suggest that our model is promising for efficient and accurate computation of exact phase diagrams of many-body lattice hamiltonians. 
\end{abstract}

\maketitle

\section{Introduction}
In condensed matter physics, 
simple many-body lattice models have been 
employed to capture the essential 
physics of intriguing phenomena like  
phase transitions, strongly correlated states, and exotic phases. 
A particularly popular and powerful simple
model is the Hubbard model, which has 
proven useful in elucidating, for instance, the phase diagram and 
the spectra of high-temperature superconductors~\cite{anderson1998theory,huang2017numerical,bedell1990high}, 
of unconventional superconductivity in 
twisted multilayered van der Waals heterostructures~\cite{wu2018hubbard,xie2019spectroscopic,pan2020band,pan2020quantum,ochi2018possible,dodaro2018phases,wong2020cascade,potasz2021exact,morales2021mit}, 
and the superfluid-to-insulator 
transition in bosonic gases~\cite{scalapino1992superfluid,freericks1994phase,kuhner1998phases,capogrosso2007phase}.
However, the solution of the deceptively simple Hubbard model remains 
a challenge due to the fact that, 
as a function of the system size and the number of particles in the system, an 
exponentially large space is required to fully describe the many-body wavefunction.

In this work, we adopt a variational approach by considering 
a class of wavefunctions parameterized by a deep neural network (DNN) 
to reduce the parameter space to search for the optimal solution. 
Similar machine learning (ML)-based variational approaches have proven 
useful in providing insights into the ground state of interacting spin system~\cite{carleo2017,nomura2017restricted,bravyi2019approximation}, 
bosonic systems~\cite{saito2017,saito2018machine,mcbrian2019ground}, 
fermionic systems~\cite{choo2020fermionic,robledo2021fermionic,yoshioka2021solving,bennewitz2021neural,robledo2021fermionic,yoshioka2021solving}, dynamics in many-body systems~\cite{hartmann2019dynamics}, and open quantum systems~\cite{nagy2019variational,vicentini2019variational}.
While the amount of information encoded 
in these ML-based models in principle scales linearly as a function of the number of particles and sites, 
the actual training of a DNN is known to 
be computationally costly and in many cases 
more costly than the exact diagonalization (ED) of the hamiltonian, 
because it requires a new training for every different value of the parameters on which the hamiltonian depends. 

The very large size of the space needed to represent the wavefunction has so far 
restricted the search for solutions, 
typically, only to the ground state of
the many-body system. 
Our proposed model 
(in the following referred to as the \texttt{HubbardNet}) 
employs a variational approach 
to solve for the ground state 
as well as for the excited states 
of the one-dimensional (1D) and two-dimensional (2D) Bose-Hubbard model. 
Our DNN-parametrized solutions are an analytic function of the on-site 
Coulomb repulsion between electrons,
described by the parameter $U$,  
and the total number of particles $N$.
This bypasses the need to perform 
an optimization for every 
value of the $(U,N)$ parameters of the 
hamiltonian.
Our results provide the 
proof of principle that 
variational approach employing DNN's
can be used to efficiently map out 
the phase boundaries in $(U,N)$ space 
of low-dimensional lattice models.

The paper is organized as follows. 
In Section~\ref{sec:model}
we introduce the Bose-Hubbard model and its construction. 
In Section~\ref{sec:dnn}
we present the details of \texttt{HubbardNet} 
and use it to solve for the 
ground state and excited states of 
the 1D and 2D Bose-Hubbard model 
on a square lattice with $M$ sites, 
using both periodic and open boundary conditions.
We evaluate the performance of our method and compare to ED in Section~\ref{sec:performance}. 
Finally, we summarize our findings in Section~\ref{sec:discussion} and discuss the possible extensions to fermionic systems.

\section{The Bose-Hubbard Model}~\label{sec:model}
The Bose-Hubbard Hamiltonian is given by
\begin{equation}
\hat{\mathcal{H}} = -t \sum_{\langle ij \rangle} \hat{a}_i \hat{a}_j^\dagger 
+ \frac{U}{2} \sum_{i=1}^M
\hat{n}_i (\hat{n}_i-1) + \sum_i^M (V_i - \mu) n_i,
\label{eqn:hamiltonian}
\end{equation}
where the sum on $i$
runs over the $M$ lattice sites 
and $\langle ij \rangle$
denotes the nearest neighbor
pairs in the lattice. 
The first term in the hamiltonian 
represents the kinetic energy of the quantum particles,
with $\hat{a}_i^\dagger, \hat{a}_i$
being the creation and annihilation 
operators for a particle at the lattice 
site $i$, 
and $-t$ being the amount of energy gained 
by a particle during a ``hop'' between 
two neighboring sites $\langle ij \rangle$.
The next term in the hamiltonian
represents the interaction between 
particles, assumed to be charged 
(electrons or ions). 
We assume the system consists of only one kind 
of charged particles, 
with $U$ describing the Coulomb repulsion 
between each pair; 
$\hat{n} =  \hat{a}_i^\dagger \hat{a}_i$ is the number operator, whose expectation 
value gives the occupation of 
the lattice site $i$. 
The last term includes a contribution from an external potential $V_i$ at site $i$, and the chemical potential energy of the system, $\mu$, which we will assume to be zero in this work. 
In the following we take $t=1$ and 
express all values of energy in units of 
$t$. 
This simple model can capture the 
superfluid phase for $U/t \leq 1$ 
and the Mott insulator phase 
for $U/t \gg 1$~\cite{potasz2021exact}.

We will denote a state, referred to as a Fock state, of the system
as $|\vec{n}_j\rangle$, which is  
defined by the occupation of each 
site $n_i^{(j)}$, with the index $i$ 
running over all $M$ sites, that is,
\begin{equation}
|\vec{n}_j\rangle = 
| n_1^{(j)}, n_2^{(j)}, \dots, n_M^{(j)}
\rangle,
\label{eq:Fock_states}
\end{equation}
with $n_i^{(j)}\in[0, N]$ for $i=1,\dots,M$, which satisfies
\begin{equation}
\sum_{i=1}^M n_i^{(j)} = N.
\label{eq:sum_occupations}
\end{equation}
The number $\mathcal{N}_B$ 
of these Fock states 
for the Bose-Hubbard model with $N$ particles in $M$ sites 
is given by 
\begin{equation}
\mathcal{N}_B = {M + N - 1 \choose N}.\label{eq:number_states}
\end{equation}
Using this basis of Fock states, the 
many-body wavefunction $|\Psi_k\rangle $ 
that describes the entire system is written as:
\begin{equation}
|\Psi_k\rangle  
= \sum_{j=1}^{\mathcal{N}_B}
\psi_j^{(k)} |\vec{n}_j \rangle,
\end{equation}
and the solution involves finding all the many-body wavefunctions that satisfy the eigenvalue equation 
\begin{equation}
\hat{\mathcal{H}} |\Psi_k\rangle 
= E_k |\Psi_k \rangle,
\end{equation}
with $E_k$ being the $k$-th energy eigenvalues. 
The states $|\Psi_k\rangle $ 
form an orthonormal set, 
$\langle \Psi_i | \Psi_j \rangle = \delta_{ij}$, 
which implies 
\begin{equation}
\sum_{j=1}^{\mathcal{N}_B} 
|\psi_j^{(k)}|^2 = 1
\label{eq:normalization}
\end{equation}
for each value of $k$.
Thus, the solution involves finding the complex numbers $\psi_j^{(k)}, 
j = 1, \dots, \mathcal{N}_B$, that describe each of the 
many-body states. 

The solution is found by constructing 
the hamiltonian matrix 
\begin{equation}
\mathcal{H}_{ij} = 
\langle \vec{n}_i |\hat{\mathcal{H}} 
| \vec{n}_j \rangle, \; \; 
i,j = 1, \dots, \mathcal{N}_B, 
\label{eq:hamiltonian_matrix}
\end{equation}
and finding all its eigenvalues and eigenvectors. 
To this end, every Fock state needs 
to be associated with an integer label, 
in a unique and known ordering. 
This can be done iteratively by Ponomarev ordering~\cite{ponomarev2009,ponomarev2010},  
a procedure described in detail 
by Revent\'os {\it et al.}~\cite{raventos2017}. 
We note that the kinetic-energy 
term produces contributions only to the 
off-diagonal hamiltonian matrix elements, 
because it involves hops of particles 
between nearest-neighbor sites, which changes 
the occupations of these sites. 
For example, for a model with $M = 5, N = 3$, supposing the $i^{\mathrm{th}}$ state 
is $\vec{n}_i = [2, 1, 0, 0, 0]$ 
and the $j^{\mathrm{th}}$ state is $\vec{n}_j = [1, 2, 0, 0, 0]$ by Ponomarev ordering, $\mathcal{H}_{ij} = -t$. 
This is because a particle at site 1 in state $\vec{n}_i$ can hop to site 2, which results in state $\vec{n}_j$. This nearest neighbor hop contributes a term of $-t$ 
to the energy.
By similar arguments, the 
Coulomb repulsion term produces 
contributions to the diagonal matrix 
elements only, because it does not involve 
any changes in the site occupations. 

\begin{figure}[ht!]
    \centering
    \includegraphics[width=\linewidth]{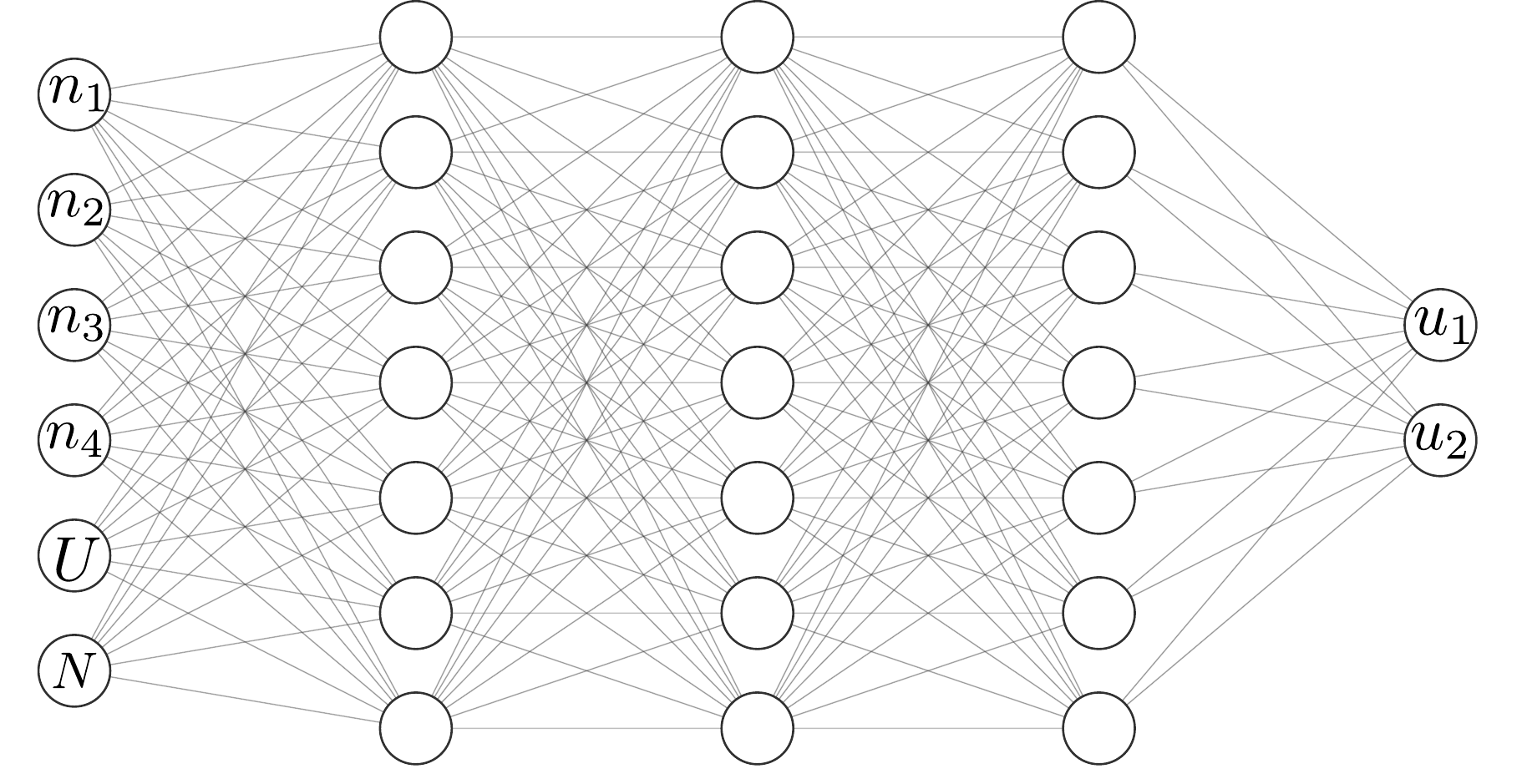}
    \caption{
    Schematic diagram of the data-free neural network to solve for the Bose-Hubbard model. The network takes $(M+2)$ inputs, which are the number of particles $n_i$ at all $i=1,\dots,M$ sites as well as $U$, the Coulomb repulsion parameter,
    and $N$, the total number of particles in the system (in the example shown we use $M=4$ for illustration purposes). The output layer is two-dimensional, with the outputs $u_1$ and $u_2$ corresponding to the real and imaginary part of the wavefunction component corresponding to this particular configuration of particles. }
    \label{fig:network}
\end{figure}

\section{Neural Network-Parametrized Solution} ~\label{sec:dnn}
We parametrize the many-body 
wavefunction $|\Psi_k\rangle$ by a fully connected DNN 
\begin{equation}
|\Psi_k \rangle = \sum_{j=1}^{\mathcal{N}_B}
g(\vec{n}_j,U,N;\vec{W}^{(k)}) 
|\vec{n}_j \rangle 
\label{eq:network_representation}
\end{equation} 
where $\vec{W^}{(k)}$ denotes the network parameters for this state; 
a schematic representation of the network is shown in Fig.~\ref{fig:network}. 
Thus,
the network takes as input 
the pair of values $(U,N)$ and
the set of occupation numbers
$n_i^{(j)}, i=1,\dots,M$, that 
correspond to a specific Fock state 
$|\vec{n}_j \rangle $, 
and gives as output the coefficient 
$\psi_j^{(k)}$ that corresponds 
to this Fock state, which is a component of the 
many-body wavefunction $|\Psi_k \rangle$
with energy $E_k$.
There is apparently a redundancy in 
these definitions, because of the constraint given in
Eq.~\eqref{eq:sum_occupations}, 
so not all of the input values to the DNN
are independent. 
However, a better description is 
that for a given value of $N$, 
only the Fock states within the 
space defined by Eqs.~\eqref{eq:Fock_states} 
and ~\eqref{eq:sum_occupations}
are allowed as inputs to the network. 
In this sense, $N$, which is an important 
variable in the problem, can be 
considered as an independent variable.
To obtain the full wavefunction 
$|\Psi_k\rangle$, the trained network 
needs to be applied to each of the  
Fock states once. 

The size of the hamiltonian matrix, $\mathcal{N}_B\times \mathcal{N}_B$, increases 
as the factorials of
the number of particles $N$ and sites $M$. The exact diagonalization of the hamiltonian matrix 
quickly becomes intractable as the system size increases. 
In contrast, the number of input parameters scales linearly with $M$ in the DNN approach~\cite{carleo2017}.
Another important advantage of the DNN-parameterized solution is that with $U$ and $N$ being the input parameters to the network, the network becomes an analytic function of $U$ and $N$ and we essentially solve for a class of hamiltonians during a single training. 
The network can then be used to make predictions at different 
values of $U$ that are not in the training set. In contrast, other techniques including ED and determinant quantum Monte Carlo~\cite{dqmc1981} require 
a new calculation for every different 
set of values $(U,N)$.  

In this work, we use 4 hidden layers and use a hyperbolic tangent, $\tanh()$, activation function for all hidden layers. 
In the examples that we considered in this work, the wavefunctions are real. 
However, the wavefunctions can in general be complex, and thus we need two real-number outputs 
for the real and imaginary parts
of the coefficients 
\begin{equation} 
\psi_j^{(k)} 
= u_{1,j}^{(k)} + iu_{2,j}^{(k)}.
\label{eq:psi_linear}
\end{equation} 
For the ground state wavefunction, 
$|\Psi_0\rangle$ 
(denoted by the index $0$), 
we choose the activation function of the output layer to be an exponential, 
namely 
\begin{equation} 
\psi_j^{(0)}  
= \exp\left ( u_{1,j}^{(0)} + i u_{2,j}^{(0)} \right ).
\label{eq:psi_exponential}
\end{equation}
This choice is made for better 
convergence with uniformly 
sampled initial weights. 
Note that with the exponential function, the wavefunction $\psi_j (\vec{n})$ is positive. 
The coefficients of the 
Fock states have the same sign for the ground state, which is guaranteed by the structure of the hamiltonian for $-t<0$ 
on all the off-diagonal matrix elements.  
For excited states, we take a linear activation function for the output layer, 
given in Eq.~\eqref{eq:psi_linear}, 
because 
the exponential activation 
function would no longer work, as 
the wavefunction needs to be orthogonal  
to the ground state wavefunction, 
and the coefficients $\psi_j^{(k)}$ 
must have mixed signs. 

The average of any operator 
$\hat{\mathcal{A}}$ that applies
to Fock states $|\vec{n}_j\rangle$
can be obtained through the 
expression
\begin{align}
\langle \hat{\mathcal{A}} \rangle_k 
 \equiv \langle \Psi_k | \hat{\mathcal{A}} 
| \Psi_k \rangle 
 = \sum_{i,j=1}^{\mathcal{N}_B}
\psi_i^{(k)*} \langle \vec{n}_i | 
\hat{\mathcal{A}} | \vec{n}_j \rangle 
\psi_j^{(k)}.
\label{eqn:variational0}
\end{align}
Using $\hat{\mathcal{H}}$ as 
the operator in Eq.~\eqref{eqn:variational0}
we can obtain the energy $E_k$ of the 
eigenstate $|\Psi_k\rangle$ of the 
hamiltonian. 
In the network representation of 
the many-body wavefunction, Eq.~\eqref{eq:network_representation},
the energy takes the form:
\begin{align}
E_k(U,N)= 
& \sum_{i,j=1}^{\mathcal{N}_B}
g^*(\vec{n}_i,U,N; \vec{W}^{(k)})
\mathcal{H}_{ij}
g(\vec{n}_j,U,N; \vec{W}^{(k)})
\nonumber \\ 
& \times \left [ 
\sum_{i=1}^{\mathcal{N}_B} 
|g(\vec{n}_i,U,N; \vec{W}^{(k)})|^2
\right ]^{-1},
\label{eq:energy_network}
\end{align}
where $\mathcal{H}_{ij}$ 
are the hamiltonian matrix elements 
defined in Eq.~\eqref{eq:hamiltonian_matrix}.
In the above equation 
we have indicated explicitly the 
dependence of the energy eigenvalues
on $U$ and $N$.
Accordingly, 
we choose the loss function as follows:
\begin{equation}
\mathcal{L}_k=\frac{1}{\mathcal{N}_U \mathcal{N}_N}\sum_{l=1}^{\mathcal{N}_U} \sum_{m=1}^{\mathcal{N}_N}
E_k(U_l,N_m),
\label{eqn:loss}
\end{equation}
where $E_k(U_l,N_m)$ is the expression 
defined in Eq.~\eqref{eq:energy_network},
and $\mathcal{N}_U$, $\mathcal{N}_N$ are the numbers 
of values of $U$ and $N$ included in the training, respectively.
The loss functions for 
the ground-state energy $E_0$ and 
for excited-state energies 
$E_k (k>0)$, need to 
be treated differently, 
as described in detail next.


\begin{figure*}[ht!]
    \centering
    \includegraphics[width=\linewidth]{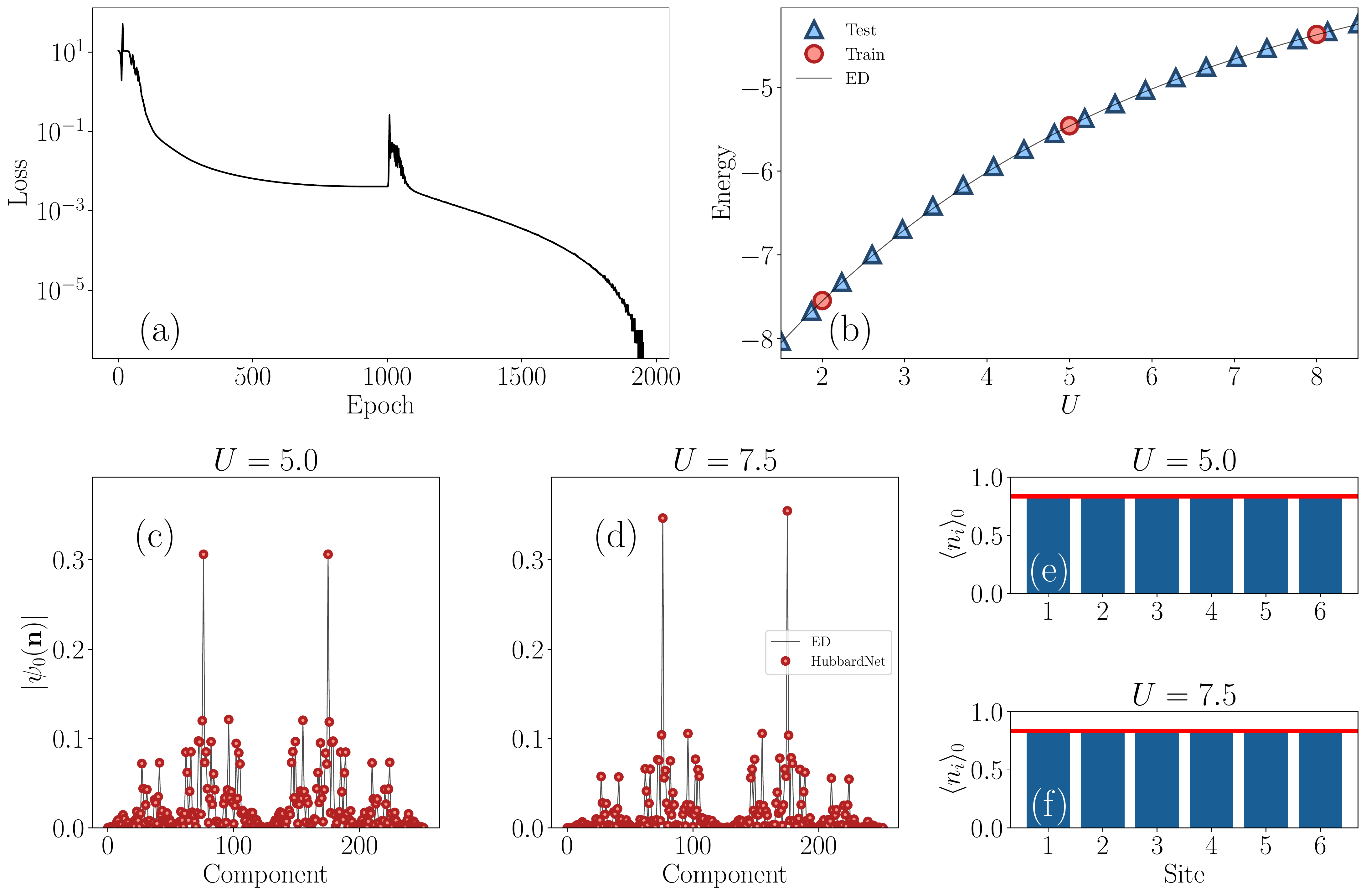}
    \caption[One-dimensional Bose-Hubbard model solution obtained using the \texttt{HubbardNet}]{One-dimensional Bose-Hubbard model solution with a periodic boundary condition and $M=6, N=5$ obtained using the \texttt{HubbardNet}. (a) Trace of the loss function offset by the negative of the minimum total energy. (b) Comparison between the ground truth ground state energies obtained from ED (black solid line) and the neural network results for training set (red scattered crosses) and prediction (blue scattered points).  (c)-(d) Absolute value of the wavefunction $|\psi_0(\vec{n})|$ from ED (black solid lines) and neural network (red scattered crosses) for (c) $U=5.0$ (d) $U=7.5$. (e)-(f) The occupation number $\langle n_i \rangle_0$ at different sites that correspond to (c)-(d). Red solid lines represent a constant occupation number $\langle n_i \rangle = N/M= 0.833$. }
    \label{fig:1d}
\end{figure*}

\subsection{Ground state}

To obtain the ground state energy and wavefunction, we minimize the loss as defined 
in Eq.~\eqref{eqn:loss} directly. 
As an example, we first discuss 
the solution for a 1D system 
with periodic boundary conditions.
In this case, every site has 2 nearest neighbors. 
We train the network with 
three different $U$ values, 
namely $U = 2.0, 5.0, 8.0$, 
for a system with $M=6, N=5$, 
which yields $\mathcal{N}_B = 252$ Fock states. 
Here and in the rest of the paper, unless otherwise specified, we choose the network width to be $D=400$. 
We use stochastic gradient descent (SGD) for the optimization of the network weights
$\vec{W}^{(0)}$ that minimize the loss function. 
We use momentum for learning $=0.9$. 
We adopt a cosine annealing scheme for the learning rate~\cite{loshchilov2016}, with the maximum learning rate chosen to be $0.01$ and decaying to zero.  
We reset the learning rate every 1000 iterations. 
We train the network until the variance of the loss function 
during the last 200 steps is less than $1\times 10^{-7}$ (see Fig.~\ref{fig:1d}(a)).

In Fig.~\ref{fig:1d}(b), we compare the
ground-state energy values 
as a function of $U$,  
as obtained by \texttt{HubbardNet}
with the values 
obtained by diagonalizing
of the hamiltonian matrix (namely ED).
The comparison includes the values of $U$ 
that are part of the training set (red scattered points), 
as well as values beyond 
the training set (blue scattered crosses).
We note that even though the energy is a smooth function of $U$, 
the output of the network includes all 
the wavefunction components 
$\psi_j^{(0)}, j=1,\dots,\mathcal{N}_B$, which are a more complex function of the 
parameter $U$.
For example, in Fig.~\ref{fig:1d}(c)-(e) we 
compare the wavefunction components 
(labeled with an arbitrary ordering to reflect the symmetry of the basis) 
from the neural network output, 
shown as a scatter plot, 
and the corresponding values 
obtained from ED, 
for three different values of $U$, 
only one of which was included in the 
training set, namely $U=5.0$ (see Fig.~\ref{fig:1d}(d)).
In all three cases, 
there is an excellent agreement between the 
two results, indicating that 
with as little as $\mathcal{N}_U = 3$ 
points in the training sets, \texttt{HubbardNet} is capable of predicting the wavefunctions and energies of the 1D Bose-Hubbard model accurately for a wide range of the values of $U$ 
In Fig.~\ref{fig:1d}(f), we show 
the average occupation number 
$\langle \hat{n}_i \rangle_0$, 
obtained from Eq.~\eqref{eqn:variational0},
using $\hat{n}_i$ as the operator, 
with the trained weights, for three different values of $U$, both within and beyond the training set. 
As expected from the periodic boundary 
conditions, and the large value of $U$, 
the average occupation number is 
uniform for all values of $U$ in this example. 
In other words, the large repulsion 
between the charged particles uniformly distributes 
them in the lattice. 

\begin{figure*}[ht!]
    \centering
    \includegraphics[width=0.9\linewidth]{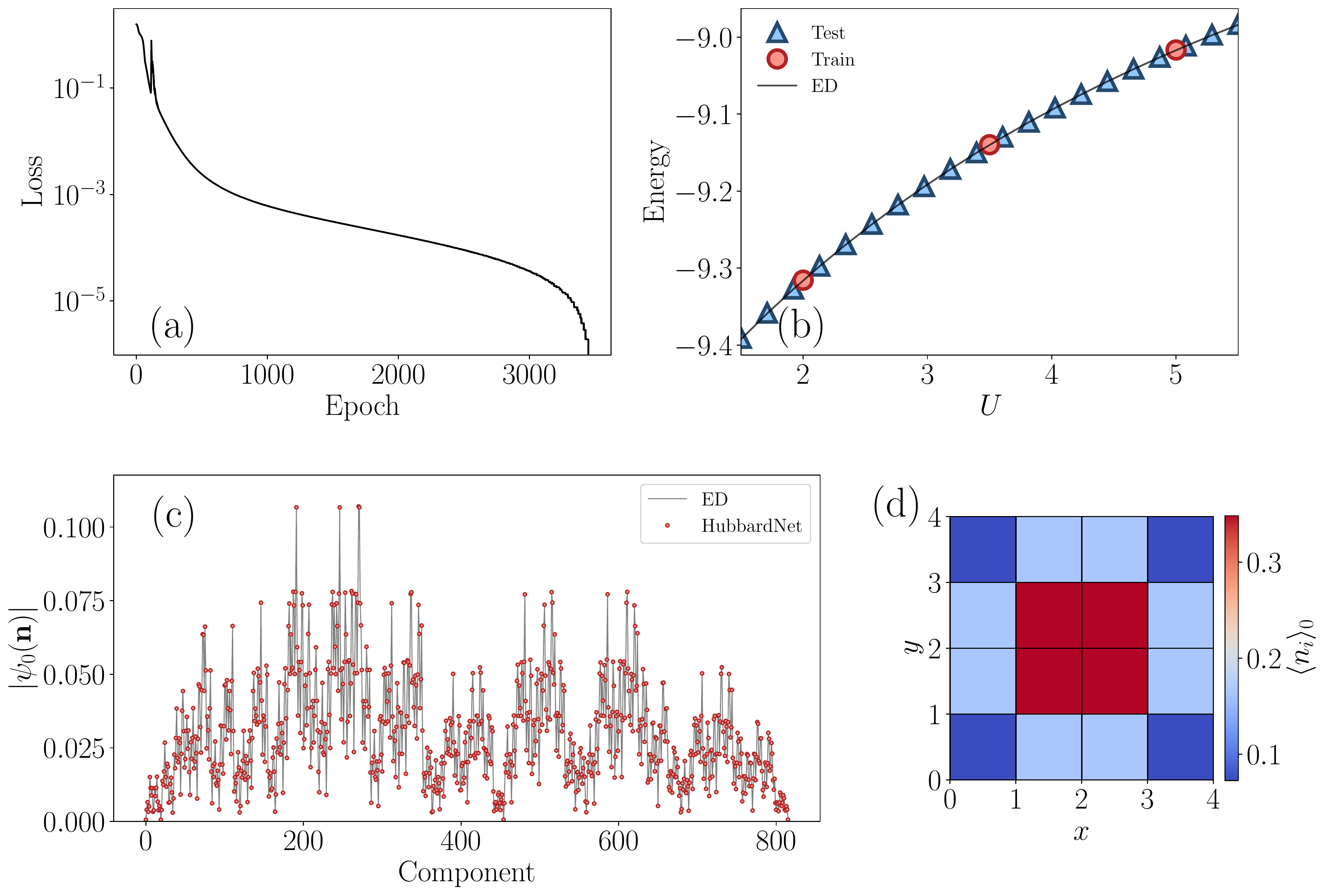}
    \caption[\texttt{HubbardNet} solution for a 2D Bose-Hubbard model]{Two-dimensional Bose-Hubbard model solution with an open boundary condition and $M=16, N=3$ with ($M_x=4, M_y = 4$) obtained using the \texttt{HubbardNet}. (a) Trace of the loss function offset by the negative of the minimum total energy. (b) Comparison between the ground truth ground state energies obtained from exact diagonalization (black solid line) and the neural network results for the training set (red scattered points) and prediction (blue scattered points). (c) Absolute value of the wavefunction $|\psi_0(\vec{n})|$ from the exact diagonalization (solid black lines) and neural network (red scattered crosses) for $U=2.5$.  (d) The average particle occupation number $\langle n_i \rangle_0$ corresponding to (c) with $U=2.5$.  }
    \label{fig:2d}
\end{figure*}

We next proceed to examine the capability of \texttt{HubbardNet} 
to handle more complicated cases, 
namely lattices with a larger number of sites and particles. 
We consider a 2D Hubbard model 
with open boundary conditions 
(a cluster of lattice sites), 
with $M=16$ (a $4\times 4$ square lattice) 
and $N=3$. These choices 
lead to a large 
number of Fock states, 
$\mathcal{N}_B = 816$. 
The hamiltonian construction is similar to the 1D case, but now 
every site has 4 nearest neighbors instead of 2, 
except for the sites on the boundary 
which have only 2 or 3 neighbors, 
from the open boundary condition. 
In this way, the hamiltonian becomes denser compared to the 1D case. 

In Fig.~\ref{fig:2d} we show 
the solution, 
using the same values of $U$ as the 
training set, namely $U = 2.0, 3.5, 5.0$. 
Despite the increased size of the problem,
\texttt{HubbardNet} performs extremely well: both the energies, shown in Fig.~\ref{fig:2d}(b), and the wavefunction components, shown in Fig.~\ref{fig:2d}(c), 
match the results from the ED solution for $U=2.5$, 
despite the fact that this value was not part of the training set. 
The occupation number distribution (Fig.~\ref{fig:2d}(d)) is no longer uniform but is instead concentrated in the center of the cluster due to the open boundary condition. 

\begin{figure*}[ht!]
    \centering
    \includegraphics[width=\linewidth]{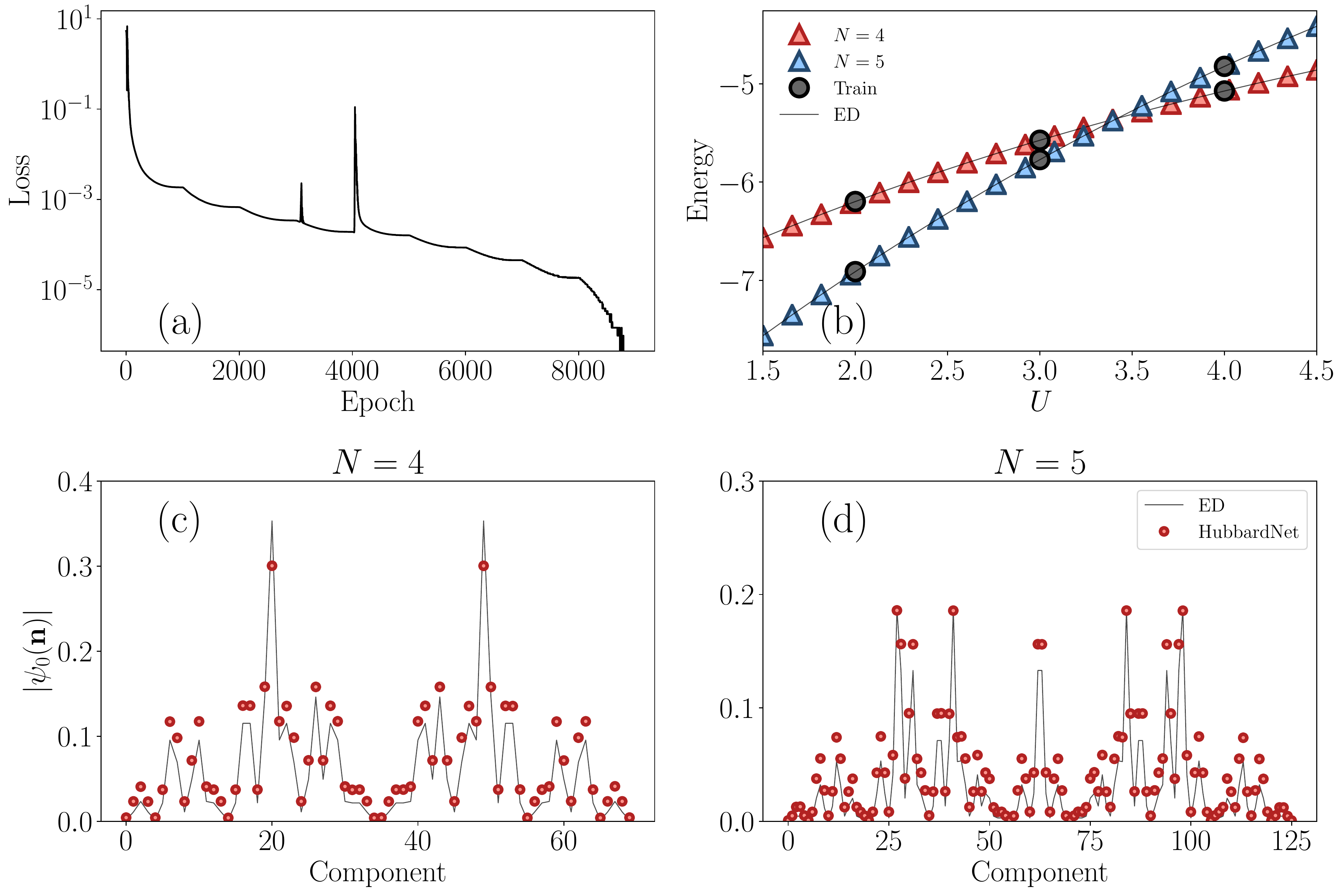}
    \caption{One-dimensional Bose-Hubbard model ground state solution with a periodic boundary condition for $M = 5$, and the training set including both the $N = 5$ and $N = 4$ states. (a) Trace of the loss function offset by the negative of the minimum total energy. (b) Energy vs. $U$ from ED (solid black lines) and \texttt{HubbardNet} (scattered points) with $N=4$ (red) and $N=5$ (blue) respectively. Black circles are from the training set. (c)-(d) Comparison between wavefunction components $|\psi_0 (\vec{n})|$ obtained from ED (blue solid lines) and \texttt{HubbardNet} (black solid lines) for (c) $N=4$ and (d) $N=5$ with $U=2.5$. } 
    \label{fig:multi_N}
\end{figure*}

In addition to including several values of  
$U$ as part of the network input, 
the number of particles in the system, $N$, 
can also be part of the network input. Although the space of Fock states 
has different size for different values of $N$, \texttt{HubbardNet} 
can still be used as long as $M$ is fixed, since the energy is an average over all particle configurations in the Fock basis. 
In Fig.~\ref{fig:multi_N}, we show the 
results from using \texttt{HubbardNet}
for a system with 
multiple values of both $N$ and $U$ as inputs, 
so that the output is a function of both variables.
Figure~\ref{fig:multi_N}(b) shows that the
two curves describing the energy as 
a function of $U$ 
for the two different values of $N$ cross
each other, making the the optimization much more difficult. 
Remarkably, energies obtained from \texttt{HubbardNet} both at the values of 
$U$ included in the training set and outside of the training set agree with the ED results (scattered points in Fig.~\ref{fig:multi_N}), despite the crossover between the $N=4$ and $N=5$ energy vs. $U$ curve. 
The \texttt{HubbardNet} wavefunction components, while retaining the approximate shape of the ED result, 
are slightly
different from the latter (Fig.~\ref{fig:multi_N}(c)-(d)). 
It is more difficult for the network to train with multiple values of $N$ in the input because each $N$ gives rise to a different Hilbert space size, and thus the corresponding wavefunctions are expected to be more different than in the case with only multiple values of $U$. 
Another difficulty is that $N$ needs to be an integer, but we are not constraining the network to predict integer values of $N$. Therefore, the parameter space for the minimization is unnecessarily large, which makes it more difficult for the network to converge. 
 
\subsection{Identifying the ground state phases}~\label{sec:pt}

The Bose-Hubbard model with on-site Coulomb and nearest-neighbor coupling interactions has two phases: the superfluid phase and the Mott insulator phase.  The phase transition is  exhibited by a commensurate number of particles $N/M \in \mathbb{N}$. For the 1D model, the phase transition is of the Berezinskii-Kosterlitz-Thouless (BKT) type~\cite{fisher1989boson}, which is very sensitive to the finite system size, making it difficult to pinpoint the phase boundary in a finite-size system. 
Since \texttt{HubbardNet} enables the efficient prediction of the wavefunctions and energy values for different $U/t$ and $N$, we can use it to identify the existence of the two phases and approximate the phase boundary. In this section, we first apply a site-dependent external potential to mimic an optical lattice and study the distribution of average particle number $\langle n_i \rangle_0$, and we then examine the finite-size scaling of the insulating gap for the 1D Bose-Hubbard model. 

\begin{figure*}[ht!]
    \centering
    \includegraphics[width=\linewidth]{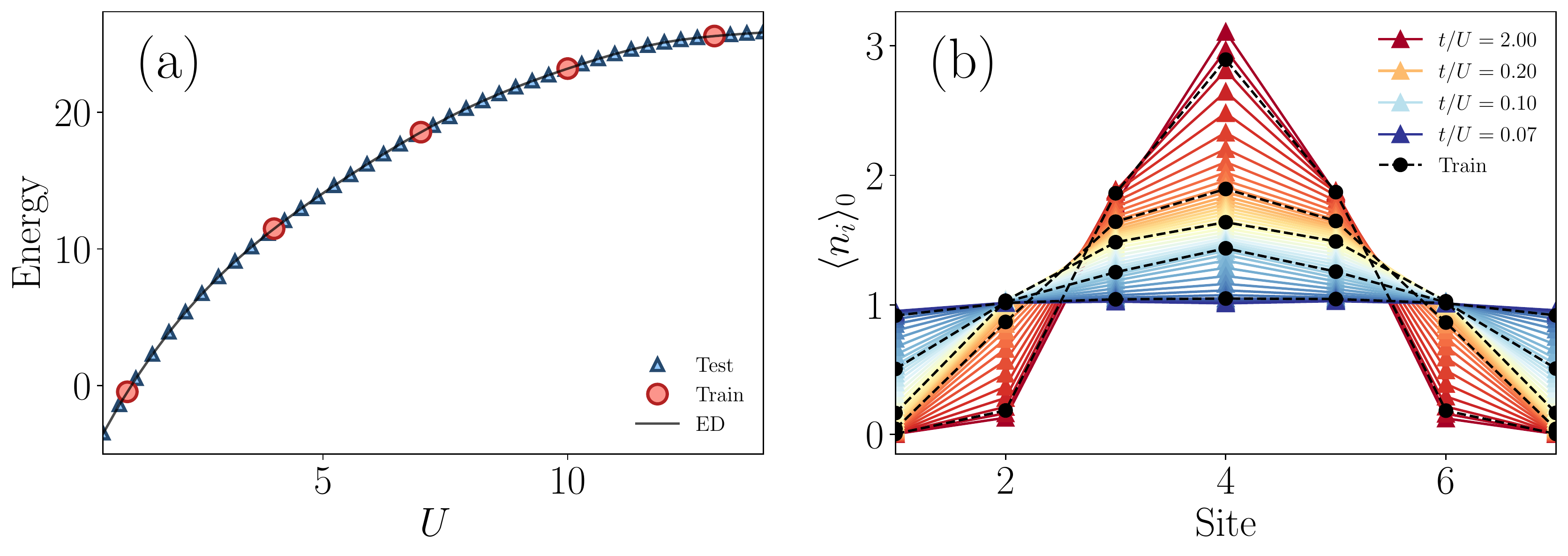}
    \caption{Energy and average particle number obtained with \texttt{HubbardNet} for $M = N = 7$ with a periodic boundary condition and a site-dependent potential with the strength $V=1$, trained with $U = 1.0,4.0,7.0,10.0,13.0$. (a) Comparison between the ground state energy \texttt{HubbardNet} (scattered points) and the ED (solid line). Red scattered points are in the training set and blue scattered points are predictions. (b) The average particle number per site $\langle n_i \rangle_0$ for different $t/U$ values. Black dashed lines are from the training set.}
    \label{fig:potential}
\end{figure*}


We choose a site-dependent external potential of the form 
\begin{equation}   
V_j = V(j - (M-1)/2)^2
\; \; for j=0,\dots, M-1  
\end{equation}
with $V=1$, 
which mimics the weak harmonic potential in ultracold atom experiments imposed due to the profile of the laser beam~\cite{bloch2005ultracold,saito2017}. 
Fig.~\ref{fig:potential}(a) shows the results of training with 5 different values of $U$ across a wide range; the energies from the training set and the testing set agree well with ED. 
Fig.~\ref{fig:potential}(b) shows the particle redistribution for different values of the $t/U$ ratio. 
At a small $U$, the particles redistribute according to the applied potential with a peak in the center of the system. At a large $U$, they distribute uniformly as the strong Coulomb repulsion prohibits them from freely hopping, which is the signature of a Mott insulator. 
We show in Fig.~\ref{fig:potential}(b) that with as little as five points in the training set, we are able to accurately predict this qualitative trend for a wide range of $U$ values.


The upper and lower boundaries of the Mott-insulator lobe are defined as the excitation energy of one particle and one hole, $\mu^\pm$ respectively.
For filling factor 
$N/M = 1$ and a fixed system size $M$, $\mu^\pm$ are given as, 
\begin{align}
\mu^+  (t/U) = E_0 (t/U, M+1) - E_0 (t/U, M), \nonumber\\
\mu^- (t/U) = E_0 (t/U, M) - E_0 (t/U, M-1). 
\end{align} 
The critical point of the Mott-superfluid transition is the value of $t/U$ at which the upper and lower boundaries of the Mott lobe coincide with each other in the thermodynamic limit~\cite{fisher1989boson}. 

\begin{figure*}[ht!]
    \centering
    \includegraphics[width=\linewidth]{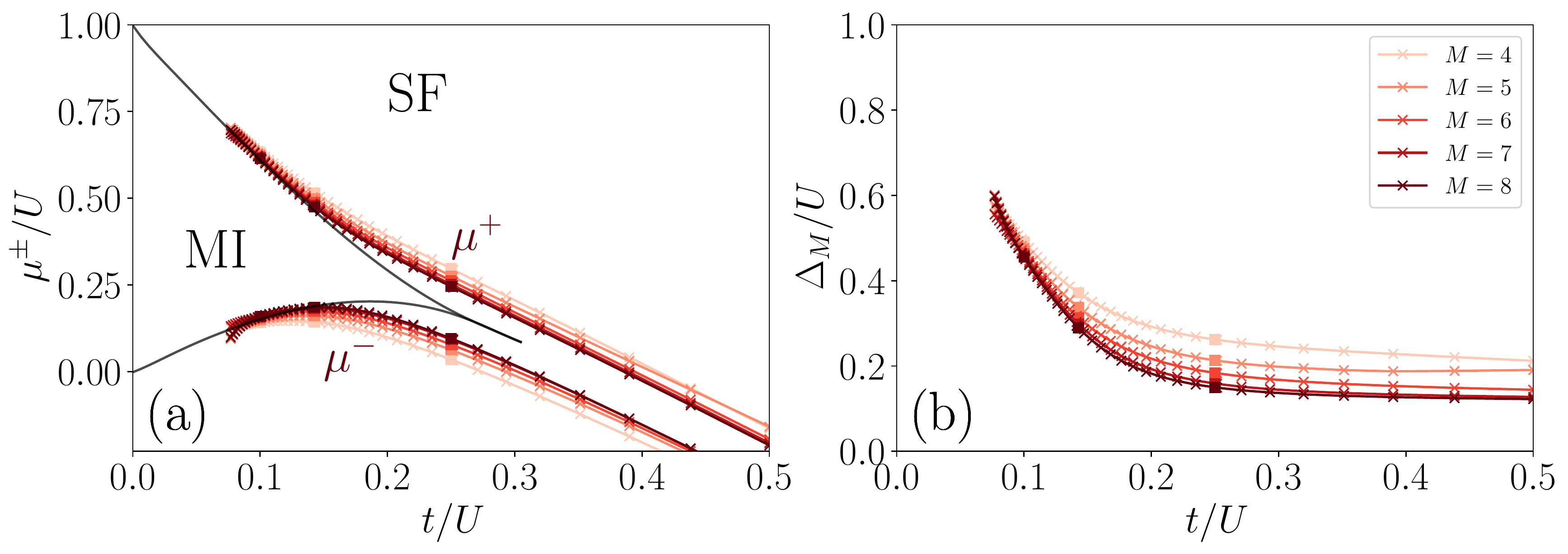}
    \caption{Mott lobe and particle-hole excitation gap obtained with \texttt{HubbardNet}. (a) The upper ($\mu^+$) and lower ($\mu^-$) boundaries of the Mott lobe for $M=N=4, \dots,  8$ obtained with \texttt{HubbardNet}. For each $M$, we train with three different values of $N$ with $N = M-1, M, M+1$ and three different values of $U$ with $U = 4.0, 7.0, 10.0$. Black solid lines are obtained using dynamical density-matrix renormalization group theory (DDMRG) with $M=128$, adapted from \citet{ejima2012characterization}. (b) Charge gap in the unit of $U$ obtained with \texttt{HubbardNet} for the same values of $M$ as (a).  }
    \label{fig:pt}
\end{figure*}

In Fig.~\ref{fig:pt}(a), we plot $\mu^{\pm}$ as a function of $t/U$ for $M=4,\dots,7$, which shows the Mott lobe and is similar to the ED result obtained in \citet{raventos2017}. 
For each $M$, we perform a single training with $N = M-1, M, M+1$ and $U = 4.0, 7.0, 10.0$. 
For a finite system size, the values of $\mu^\pm$ are not expected to coincide~\cite{raventos2017}. Therefore, we calculate $\mu^\pm$ as a function of $M$ and note that $\mu^\pm$ approach each other as $M$ gets larger. 
At a sufficiently large $M$, we expect $\mu^\pm$ to asymptotically approach the black line in Fig.~\ref{fig:pt}, which is obtained with dynamical density matrix renormalization group theory with $M=128$~\cite{ejima2012characterization}. 

The single-particle gap of the Mott insulator phase is defined as the difference between $\mu^\pm$: 
\begin{align} 
\Delta_M (t/U) &= \mu^+(t/U) - \mu^-(t/U) 
\end{align} 
Fig.~\ref{fig:pt}(b) shows $\Delta_M/U$ for the same values of $M$ as in ~\ref{fig:pt}(a); 
for increasing $t/U$, 
the gap size decreases as expected from the Mott to superfluid transition. 
In an infinite system, we expect the gap to close for the superfluid state, with the $t/U$ value at which the gap closes being
the critical point. 
We can estimate the critical point value $(t/U)_c$ 
following a similar fitting procedure as in \citet{kashurnikov1996mott}, namely, by performing
a polynomial fit to $\Delta_M/U$ as a function of $1/M$ for each value of $t/U$ and finding 
the $t/U$ value at which $\Delta_M/U$ stops varying. \citet{kashurnikov1996mott} and \cite{raventos2017} fit $\Delta_M/U$ with a degree 5 polynomial. 
Here, as a crude estimate, we perform a linear fit 
to $\Delta_M/U$ as a function of $1/M$ at each $t/U$ due to the limited number of values of $M$ at our disposal. 
In this way, $\Delta_\infty/U$ is the intercept of the fit. 
From this procedure, 
we estimate 
the critical value 
to be $(t/U)_c \approx 0.320 \pm 0.013$.
In comparison, the value 
obtained by the determinant quantum Monte Carlo (DQMC) method is $(t/U)_c = 0.275 \pm 0.005$~\cite{kashurnikov1996mott} and the value obtained with the 
DMRG method
is $0.305\pm 0.001$~\cite{ejima2011dynamic}. 
While we our value does 
not agree with these other two values
within the error bars, 
our result is only meant to provide a crude estimate with a rather limited number of data points. 

\subsection{Excited states} ~\label{sec:excited}
\begin{figure*}[ht!]
    \centering
    \includegraphics[width=\linewidth]{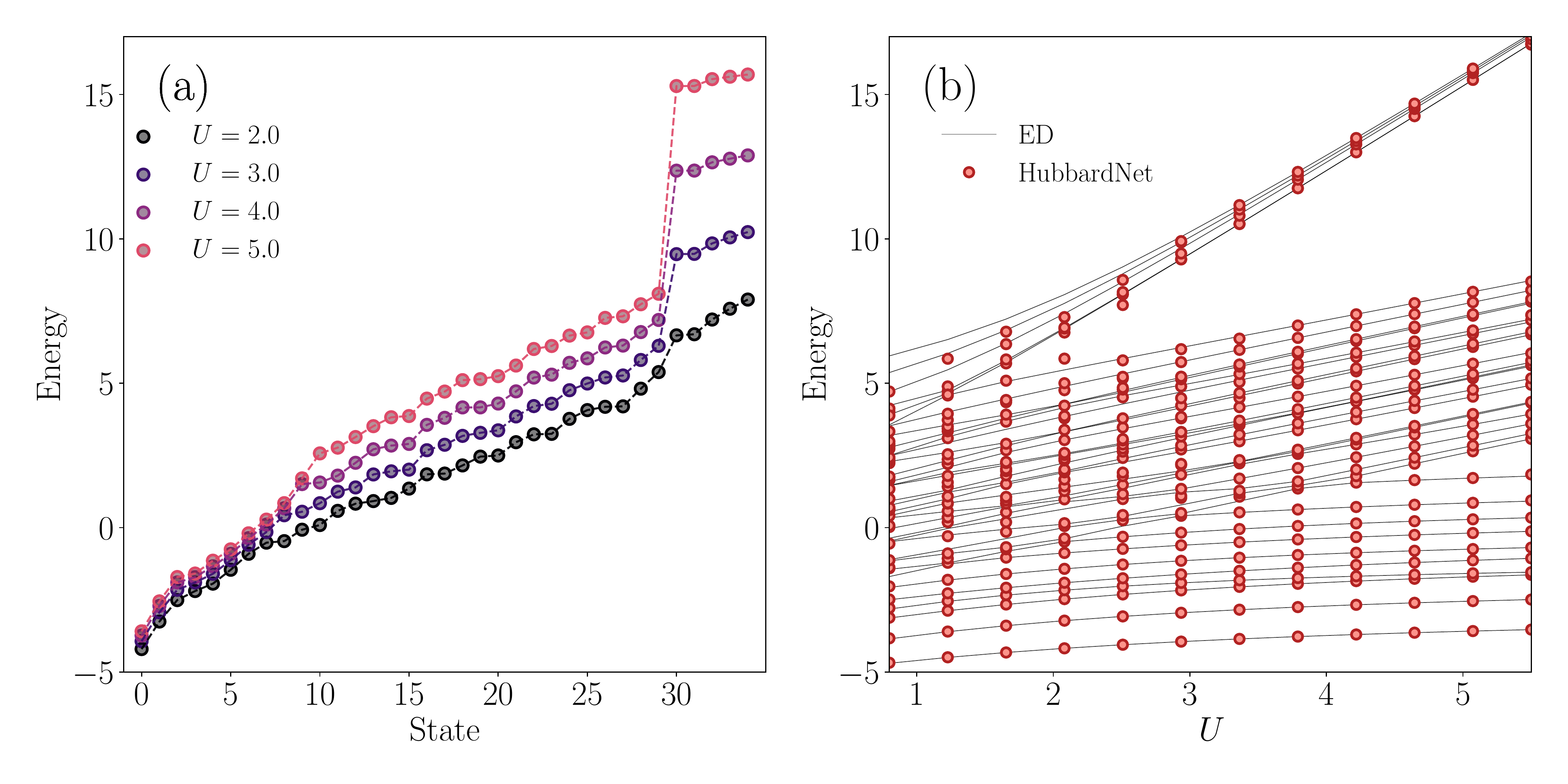}
    \caption[Energy spectrum from \texttt{HubbardNet}]{\texttt{HubbardNet} energy spectrum trained using $U=2.0,3.0,4.0,5.0$ as the input, with L2-regularization with the coefficient $=1\times10^{-4}$ and perturbing the $U$ by an amplitude $0.01$. (a) Energy from the \texttt{HubbardNet} (scattered points) vs. ED (dashed lines) for the training set (b) Energy spectrum for different values of $U$ in the testing set (scatter points). Eigenvalues from ED are in black solid lines. }
    \label{fig:spectrum}
\end{figure*}

In order to obtain excited states 
with the approach described above, we 
employ the fact that 
the eigenstates of the hamiltonian are orthogonal to each other, that is, $\langle \Psi_i | \Psi_j \rangle = \delta_{ij}$. 
An excited state is a local minimum of
the energy. 
Therefore, we can obtain excited states  
by minimizing the expectation value in Eq.~\eqref{eq:energy_network} for 
state $|\Psi_k\rangle$ with the constraint that 
it is orthogonal to $|\Psi_0\rangle, ..., |\Psi_{k-1}\rangle$.
The constraint is achieved iteratively through the Gram-Schmidt orthogonalization process.
To facilitate the calculations, we express 
the wavefunctions $|\Psi_k\rangle $
as vectors $\vec{v}_k$ of dimension 
$\mathcal{N}_B$, with entries $\psi_j^{(k)}$.
In this notation, 
assume that $\vec{v}_0, 
\vec{v}_1, ..., 
\vec{v}_{k-1}$ 
are the vectors that 
represent the output directly from the neural network, which are linearly independent, 
with $|\Psi_0\rangle = \vec{v}_0$
being the ground state wavefunction, 
and the rest being the wavefunctions 
of excited states. 
The excited state wavefunctions are obtained iteratively as follows:  
\begin{equation}
|\Psi_{k} \rangle=
 \vec{v}_k - \sum_{j=0}^{k-1} 
\frac{\vec{v}_j\cdot \vec{v}_k }{ \vec{v}_j \cdot \vec{v}_j} \vec{v}_j.
\label{eq:orthogonalization}
\end{equation}
This expression for 
state $\vec{v}_k$ incorporates the projection
out of this state of components that correspond
to all previous states, $j=0,\dots,k-1$.
The projection is performed at every iteration and the set of $\vec{v}_j$
vectors is used to compute the loss function. 
In other words, for an excited state $k$, we train a new network with the outputs being constrained to be 
orthogonal to the 
space of all states with lower energy, 
from the ground state to the $(k-1)$ 
excited state. 
With the excited state optimization
for multiple values of $U$ 
as the input, we add the following penalty term to the loss function:
\begin{align}
\mathcal{L}_P = - \exp\left ( 
- |\bar{E}_k - \bar{E}_0| \right ),
\end{align}
where the index $k$ labels the excited state, 
$\bar{E}_k$ is the average 
energy over the values of $U$ 
as obtained at each epoch and $\bar{E}_0$ is the averaged ground state energy. 
This choice of the loss function 
encourages the excited-state energy $E_k$
to be as close to the ground state energy as possible.
Similarly to the ground state, we can train the network as a function of $U$ for a given excited state $k$. 
In Fig.~\ref{fig:spectrum}(a) 
we show the energy spectra for 
4 different values of $U$, 
namely $U=2.0,3.0,4.0,5.0$, with one training for a given $k$, 
for $M=5,N=3$ with an open boundary condition.

\begin{figure}[ht!]
    \centering
    \includegraphics[width=\linewidth]{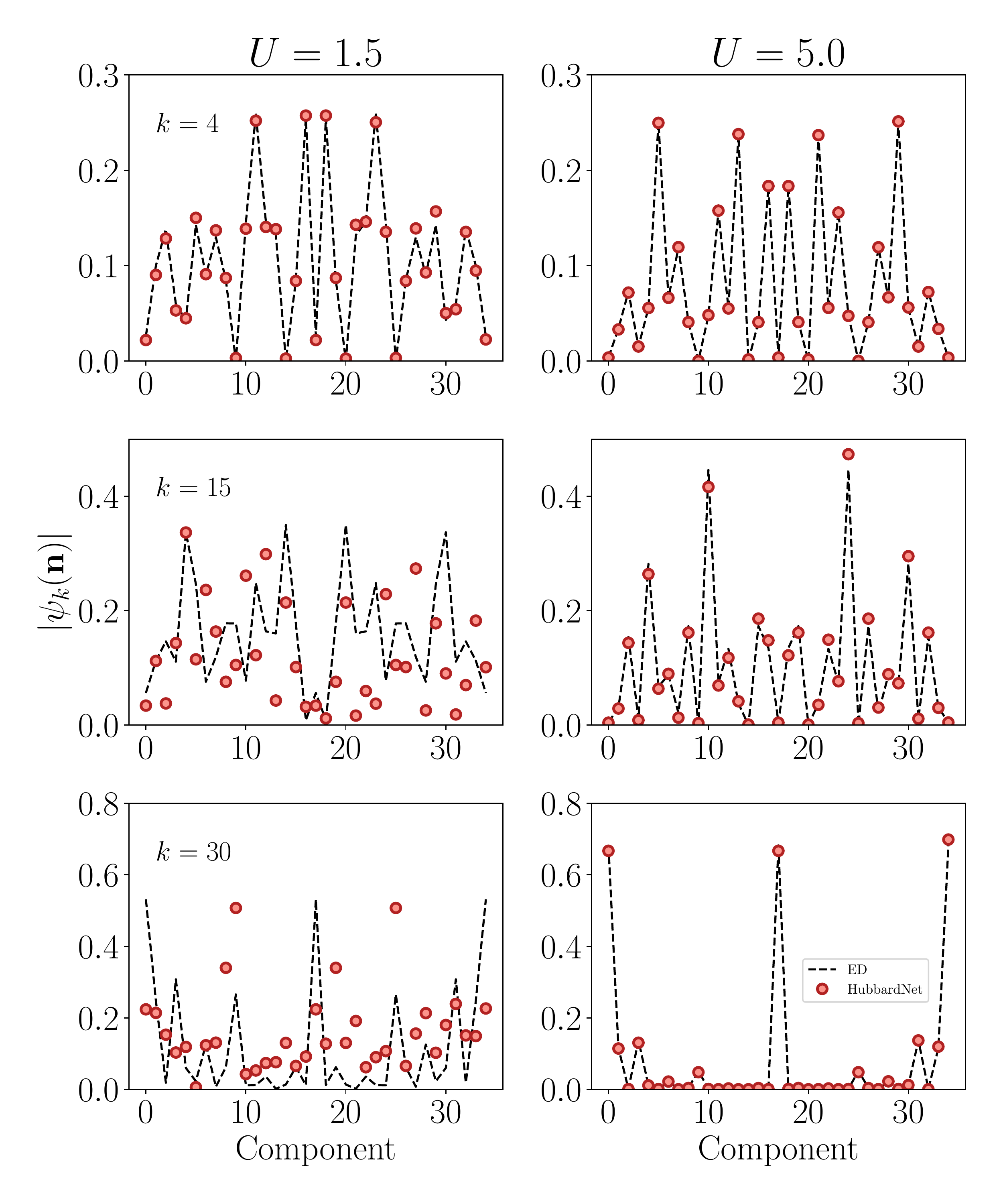}
    \caption[Excited state wavefunctions from \texttt{HubbardNet}]{Comparison between excited state wavefunctions from ED and \texttt{HubbardNet} for the $4^\mathrm{th}$ (top row), the $15^\mathrm{th}$ (middle row), and the $30^\mathrm{th}$ (bottom row) excited states for $U=1.5$ (left) and $U=5.0$ (right). In all subpanels, black dashed lines are from ED and red scattered points are from \texttt{HubbardNet}.}
    \label{fig:excited_wf}
\end{figure}

We apply an L2 regularization by adding the squared sum of the network weights to the loss function with a coefficient $10^{-4}$ in order to discourage energies from being drastically different for similar values of $U$, 
an issue that we observed in the absence of such regularization.
In addition, since our training set only contains a very small number of values of $U$ but we would like to predict energies and wavefunctions as a smooth function of all values of $U$, we randomly perturb the value of the input $U$ during every epoch, with 
a perturbation amplitude of 0.01. 
In Fig.~\ref{fig:spectrum}(a) 
we show that for the $U$ values in the training set, 
there is an excellent agreement between 
the results obtained with \texttt{HubbardNet} and those 
obtained with ED.  
Using the network, we can plot the full spectrum of the Bose-Hubbard model for the given $M$ and $N$ and an arbitrary 
value of $U$. 
In Fig.~\ref{fig:spectrum}(b) we show the spectrum produced 
using the trained network at inference. The spectrum qualitatively agrees with the ED results, 
except for $U \lesssim 2$ and at high energies. 
This is expected because $U < 2$ corresponds to values beyond the range of training-set values, 
and higher-energy states are more difficult to obtain from 
the network due to 
error accumulation in the 
iterative orthogonalization scheme, 
Eq.~\eqref{eq:orthogonalization}.

We note that there is a gap
in the energy spectrum at approximately $U=2.0$, separating a few of the high-energy excited states from the rest. 
It is also possible that the network has more difficulty 
in making accurate predictions near the phase boundary, which by definition means that a small perturbation in 
the value of $U$ leads to a large change in the qualitative behavior, and thus a denser sampling is more desirable near 
such values of $U$. 

In Fig.~\ref{fig:excited_wf}, we show the corresponding wavefunctions for three different excited states. For the results 
presented in this figure, only $U=5.0$ is included in the training set. 
For excited states with relatively low energy, for example, 
the 4$^\mathrm{th}$ excited state, the predicted wavefunctions agree with the ground truth even for values of $U$ 
not in the test set.
However, for excited states with higher energy, such as the 15$^\mathrm{th}$ or the 30$^\mathrm{th}$ state, there is a significant difference between the \texttt{HubbardNet}-predicted wavefunctions and the ground truth, especially for $U\lesssim 2$. The discrepancy in the wavefunction 
is also reflected in 
the corresponding energy spectrum in Fig.~\ref{fig:spectrum}.

As an alternative approach to the state-by-state iterative optimization,
one could treat the state $k$ as an input to the network 
and minimize the total energy of multiple states simultaneously. 
With this approach, 
we first project the neural network onto the nullspace of the ground state 
and minimize the average energy over $k$ states in the nullspace. 
In this case, the loss function landscape becomes very complex and has many degenerate local minima. 
While this approach avoids the error accumulation of the iterative optimization approach, 
we found that the network produces less accurate predictions and the results depend more sensitively on the training parameters.

\section{Performance Evaluation}~\label{sec:performance}

\begin{figure}[ht!]
    \centering
    \includegraphics[width=0.8\linewidth]{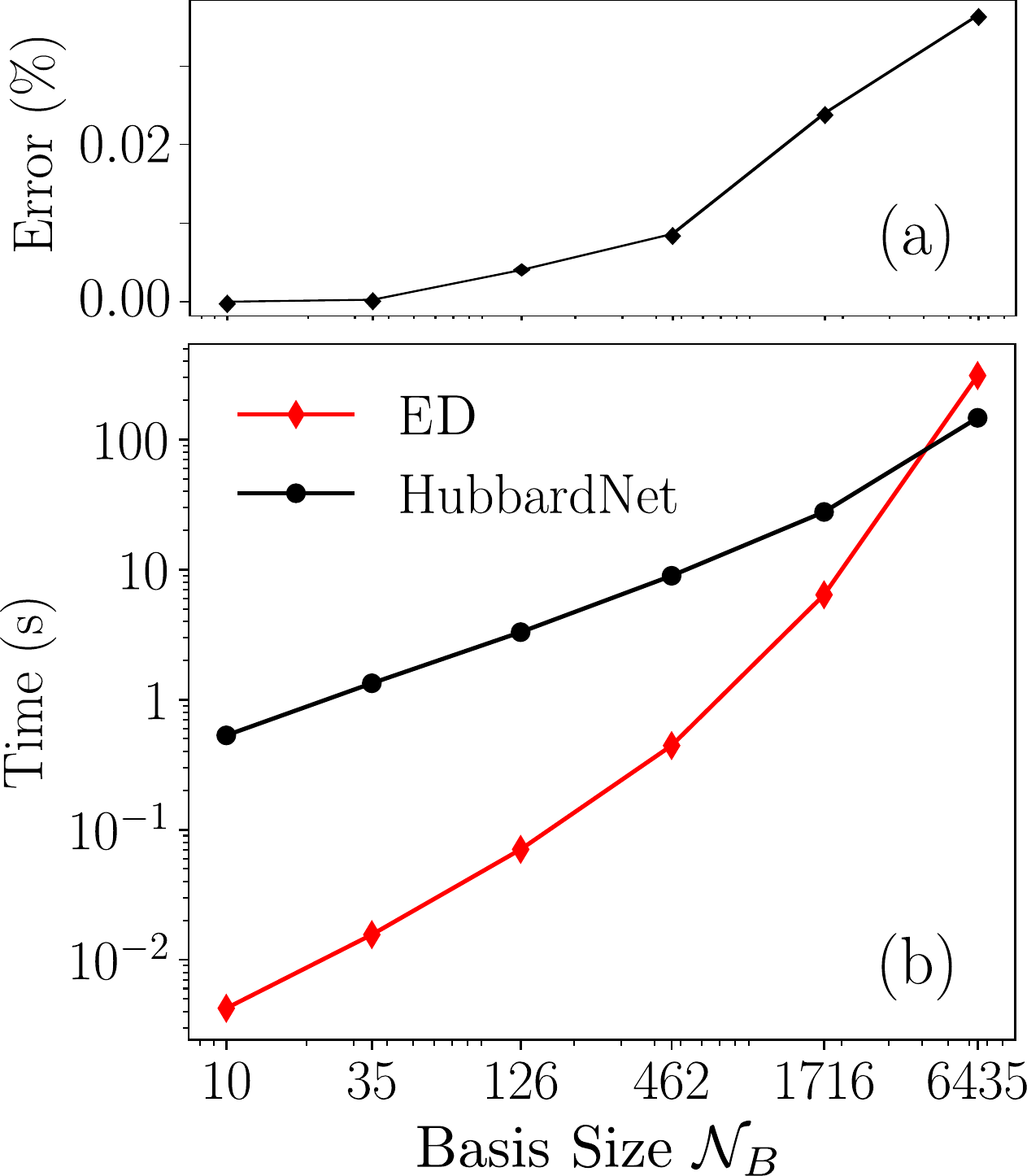}
    \caption{Performance comparison between ED and \texttt{HubbardNet} using a 1D chain (for $M=N=3,\dots,8$) and periodic boundary conditions, as a function of the system basis-size $\mathcal{N}_B$, see Eq.~\eqref{eq:number_states}. (a) Percentage error of the ground state energies obtained with \texttt{HubbardNet} versus ED (ground truth). (b) Computation time for \texttt{HubbardNet} (black) and ED (red).   }
    \label{fig:scaling}
\end{figure}

In all the simulations 
presented in Sec.~\ref{sec:dnn}, 
we used a single Nvidia K80 GPU 
and most of the calculations 
can be carried out on CPUs, 
which suggests that our approach is relatively computationally inexpensive. 
In Fig.~\ref{fig:scaling}, 
we compare the performance of obtaining 
the ground state energy for a single 
value of $U$ using 
ED and 
the \texttt{HubbardNet} method, performed on a 2.7-GHz Quad-Core Intel Core i7. 
Here, we take $M=N$ and 
a network width $D=200$. 
We adopt an optimizer constant learning rate $=0.01$ 
without L2 regularization, 
and our convergence 
criterion is the variance of the loss smaller than 
$1\times 10^{-6}$ for all cases. 
In Fig.~\ref{fig:scaling}(a), we show
that the ground state energies obtained from ED and \texttt{HubbardNet} are in excellent agreement for all cases. In Fig.~\ref{fig:scaling}(b) we show 
the computational time as a function of the system size: 
\texttt{HubbardNet} has a better scaling compared to the ED method, 
even though the latter method
is faster for smaller systems, 
with crossover being in the range 
$\mathcal{N}_B \sim 5000$.
Since the goal of numerical studies of 
the Hubbard model is to reach as large 
systems as possible, the advantage 
of the \texttt{HubbardNet} method 
is obvious. 

An alternative approach to further improve the scaling of the DNN-based
approach is that instead of calculating the sum over every pair of $\vec{n}_i, \vec{n}_j$ vectors in Eq.~\eqref{eqn:variational0}, 
one could evaluate the expectation value using a Monte Carlo sampler, which has been implemented in the package \texttt{netket}~\cite{netket2019}. 
However, due to the stochastic nature of the Monte Carlo sample, we observed that the energy to deviate from the ground truth. The accuracy of predictions 
is expected to deteriorate even more 
for excited states of higher energy 
due to the error accumulation in the iterative scheme.

\section{Discussion and Outlook}
\label{sec:discussion}

In addition to the scaling advantage
discussed in the previous section, 
our approach provides the analytical solution 
for a family of hamiltonians 
with different $U$ and $N$ values 
without the need to diagonalize a new hamiltonian with every change of 
each parameter. 
This suggests that after one training
over a set of values of $U$ and $N$, 
the network could 
in principle predict the phase diagram efficiently. 
We used \texttt{HubbardNet} to show that it is capable of identifying the ground state phases 
and provides the correct finite-size scaling for the Mott insulating gap
(Sec.~\ref{sec:pt} ). 
The ability to accurately locate the phase transition point is still limited by the training time,
which can be improved by implementing more efficient architectures such as a convolutional neural networks~\cite{saito2018machine} and by using a Monte Carlo sampler~\cite{carleo2017,saito2017,netket2019}. 
We provide a proof-of-principle demonstration in Fig.~\ref{fig:spectrum}(b) that after training on a very small training set, 
we were able to obtain the energy spectrum of the system as a function of $U$. 
In particular, the network is able to predict the gap opening near $U=2.0$. 
As we point out in Sec.~\ref{sec:excited}, energies and wavefunctions from \texttt{HubbardNet} deviate at the gap opening point, and we suggest that this problem can be solved 
by increasing the training set size. 
The choice of training set size is a tradeoff between accuracy and training time: 
the larger the training set, the more accurate the predictions should be, but the harder it is to converge and the more effort is needed to tune the 
values of the hyperparameters. 
We also note that while the scaling of \texttt{HubbardNet} is promising, the performance evaluation we showed is only for a single state. 
The task of obtaining all 
excited states is significantly more computationally costly 
due to the iterative nature of our approach. 
Nevertheless, for most practical purposes, it is sufficient to compute only the lowest few energy states, in which case \texttt{HubbardNet} would be an appropriate choice. 
Obtaining the first few excited states allows us to compute the finite temperature properties and identify the critical temperature. For example, for a given temperature $T$, we can compute an observable by the Boltzmann distribution: 
\begin{equation}
\hat{\mathcal{O} } (\beta) = \frac{1}{\mathcal{Z}} \sum_k \langle \Psi_k | \hat{\mathcal{O}} | \Psi_k \rangle e^{-\beta (E_k  - E_0)}, 
\end{equation} 
where $\beta=1/k_BT$, $k_B$ is the Boltzmann constant, $E_k$ is the energy of the $k$-th excite state, and $\mathcal{Z} = \sum_k \exp[-\beta (E_k-E_0) ]$ is the partition function. Computing the temperature dependence of observables and extracting the critical temperatures from \texttt{HubbardNet} merits future work. 

While we only provided here 
a proof-of-concept demonstration of \texttt{HubbardNet}, 
this approach could be easily adapted to solve
more realistic problems including longer-ranged hopping parametrized according to the specific materials' properties as well as systems in the presence of an external potential. 
In addition, in this work, we solve the Bose-Hubbard model in real space and obtained the wavefunction components in the Fock basis. 
We could perform a Bloch expansion of Eq.~\eqref{eqn:hamiltonian} 
in momentum space and variationally minimize the expectation of the hamiltonian, 
with the momentum $\vec{k}$ 
taken as an input to the network, which enables very fast prediction of the band structure.

An interesting extension of the \texttt{HubbardNet} approach would be to use it for the study of fermionic systems and 
for obtaining the phase diagrams of low-dimensional quantum systems including twisted bilayer van der Waals heterostructures and high-temperature cuprate superconductors~\cite{cao2018unconventional,fradkin2015highTc,heidrich2004comment}. 
The advantage of \texttt{HubbardNet} is that the fermionic wavefunction symmetry can be directly incorporated into the model hamiltonian and the basis set construction, similar to ED. 
For example, 
in twisted bilayer transition metal dichalcogenides (TMDCs),
where unconventional superconductivity 
has been reported~\cite{wang2020correlated}, the Coulomb interaction $U$ strength is related to the band flatness 
and $t$ is related to the bandwidth, both of which are functions of 
the twist angle $\theta$. 
It has been shown both theoretically~\cite{wu2018hubbard,angeli2022janus} and experimentally~\cite{xu2022tunable} that twisted bilayer TMD can be accurately represented with a Fermi-Hubbard model on a triangular, honeycomb, or Kagome lattice.  
From the results of the first-principles
calculations, one could derive 
reduced-band models to obtain the $U/t$ ratio as a function of $\theta$. 
\texttt{HubbardNet} essentially allows for the mapping of the phase boundary in twisted TMDCs as a function of $\theta$ with a single training. 
Our code is openly available~\cite{zhu2022github}.

\acknowledgements{ Z.Z. and E.K. acknowledge the support from the STC Center for Integrated Quantum Materials, NSF Grant No. DMR-1231319, and NSF DMREF Grant No. 1922172, and a Simons Foundation Award No. 896626. Calculations were performed on the Cannon cluster supported by the FAS Division of Science, Research Computing Group at Harvard University, Google Colaboratory, and the Sherlock Cluster supported by the Stanford Research Computing Center.}

\bibliography{reference}

\end{document}